\documentclass[journal]{IEEEtran}

\usepackage{cite}
\usepackage{url}
\usepackage{graphicx}
\usepackage{color}
\usepackage{float}
\usepackage{tabularx,colortbl}
\usepackage{ifthen}
\usepackage[greek,english]{babel}

\usepackage{amsmath}
\usepackage{amsthm,amsfonts}
\usepackage{amssymb}
\usepackage{pgfplots}

\usepackage{tikz}
\usepackage[customcolors]{hf-tikz}
\usepgfplotslibrary{colorbrewer}
\usetikzlibrary{mindmap,trees}
\usetikzlibrary{%
  3d,
  arrows,%
  shapes.misc,% wg. rounded rectangle
  shapes.arrows,%
  chains,%
  calc,%
  matrix,%
	fit,%
  positioning,% wg. " of "
  scopes,%
  decorations.pathmorphing,
  shadows,%
  decorations.markings,
  shapes.geometric,
  arrows.meta,bending,
	patterns,
	intersections, 
	pgfplots.fillbetween
}

\usepackage{tikz-3dplot}
\usepackage[pdfusetitle]{hyperref}
\hypersetup{
    colorlinks,
    linkcolor={red!50!black}, 
    citecolor={blue!50!black},
    urlcolor={blue!50!black}
}

\usepackage{cellspace}
\newcommand{\gpi}{\textrm{\greektext p}}
\renewcommand{\pi}{\gpi}

\providecommand*{\M}[1]{\mathbf#1}

\providecommand*{\mrm}[1]{\mathrm{#1}}

\providecommand*{\V}[1]{\boldsymbol#1}
\providecommand*{\UV}[1]{\hat{\boldsymbol#1}}
\providecommand*{\wUV}[1]{\widehat{\boldsymbol#1}}
\providecommand*{\T}[1]{\mathrm{#1}}
\DeclareMathAccent{\ring}{\mathalpha}{operators}{"17}

\providecommand*{\diffS}{\operatorname{dS}\!}

\providecommand*{\svd}{\operatorname{svd}}

\newcommand{\Na}{N_\mrm{a}} % shadow area
\newcommand{\Np}{N_\mrm{p}} % paraxial
\newcommand{\regT}{\varOmega_\T{T}}

\newcommand{\regR}{\varOmega_\T{R}}

\newcommand{\sigmat}{\tilde\sigma}

\colorlet{dpurple}{blue!50!red}
\colorlet{dblue}{blue!50!black}
\colorlet{dgreen}{green!50!black}
\colorlet{dred}{red!50!black}
\colorlet{dyellow}{yellow!50!black}
\colorlet{dorange}{orange!50!black}
\definecolor{metal}{RGB}{218,165,32}
\definecolor{diel}{RGB}{1,165,32}
\definecolor{antenna}{RGB}{100,150,162}
\definecolor{breg}{rgb}{0.2,0.6,0.8}%
\definecolor{preg}{rgb}{0.8,0.2,0.2}%
\definecolor{reg}{RGB}{218,165,32}
\colorlet{treg}{blue!50!white}
\colorlet{rreg}{red!50!white}
\tikzset{>=latex}

\makeatother

\title{From Multimode Near-Field Coupling to Friis}

\author{Mats Gustafsson%, 
\thanks{This work was supported by the Swedish Research Council (SEE-6GIA).}%
\thanks{M. Gustafsson is with Lund University, Lund, Sweden, (e-mails: mats.gustafsson@eit.lth.se).}
}

\begin{document}

\maketitle

\begin{abstract}
Near-field propagation between finite apertures can support multiple spatial channels, while far-field transmission is effectively single mode and follows the Friis transmission formula. This letter establishes a direct connection between these two regimes through the mutual shadow area between the transmitting and receiving apertures. The mutual shadow determines the asymptotic number of spatial degrees of freedom and, in the paraxial regime, leads to a simple beamforming distance that provides a characteristic scale for the transition from multimode to effectively single-mode propagation. The channel singular-value sum rule further shows that, below this distance, increasing separation primarily reduces the number of supported spatial channels while their average normalized strength remains approximately constant. Beyond the transition, the remaining channel strength decreases according to the conventional free-space quadratic power law, recovering the Friis transmission formula. Numerical results for circular apertures demonstrate the transition, the scaling with aperture size, and the dependence on the transmitter and receiver apertures. The mutual-shadow formulation also extends the description beyond the paraxial regime.
\end{abstract}

\section{Introduction}
\IEEEPARstart{N}{ear-field} beamforming is becoming increasingly important in antenna,
imaging, and wireless communication systems, particularly with the emergence
of electrically large apertures and large-scale antenna arrays
\cite{Liu+etal2023,Dardari2020}. In the far field, antenna gain and effective
area, together with the Friis transmission formula
\cite{Schelkunoff+Friis1952,Balanis2005}, provide simple and physically intuitive
measures for describing antenna links. A similarly simple description of
coupling between finite apertures in the near field is less established,
since both the number and the strength of the supported spatial channels
depend on the aperture geometries and their separation.

A distinctive feature of near-field propagation is the possibility of
supporting multiple independent spatial channels. Such spatial modes and the corresponding number of spatial degrees of freedom (NDoF) have been extensively studied for finite source and observation regions
\cite{Miller2019,Yuan+etal2022,Kaushik+DiRenzo2027}, large intelligent surfaces
\cite{Dardari2020,Decarli+Dardari2021,Li+etal2026}, and near-field MIMO systems
\cite{Xie+etal2023,Ruiz-Sicilia+etal2024}. In the paraxial regime, simple
expressions relate the spatial DoF to aperture dimensions, wavelength, and
separation distance. More recently, effective DoF measures and spatial multiplexing
capability have also been used to define communication-oriented near- and
far-field boundaries
\cite{Sun+etal2025,Song+etal2025,Hussain+etal2026}. These criteria
complement the conventional Rayleigh or Fraunhofer distance, which primarily
quantifies the validity of a planar-wavefront approximation rather than the
number of spatial channels supported by a link between two finite apertures.

Increasing the separation between two finite apertures affects their
electromagnetic coupling in two distinct ways. In a multimode
regime, higher-order spatial channels progressively disappear with increasing
separation. Once the channel becomes effectively single-mode, further
separation can no longer reduce the number of available channels and instead
reduces the strength of the remaining mode according to the quadratic
free-space power decay described by the Friis transmission formula. This
suggests that spatial DoF and far-field power transfer can be viewed as two
limits of the same aperture-to-aperture propagation problem.

In this letter, we show that the mutual shadow area provides a geometrical
description of this transition~\cite{Gustafsson2025c}. The mutual shadow determines the asymptotic
NDoF for arbitrarily shaped and oriented apertures and reduces in the
paraxial regime to
$N_\T{p}=A_\T{T}A_\T{R}/(\lambda^2d^2)$, where $A_\T{T}$ and
$A_\T{R}$ are the projected aperture areas, $d$ is their separation, and $\lambda$ is the wavelength. Combining this expression
with the channel singular values sum rule~\cite{Miller2019} reveals two complementary interpretations of $N_\T{p}$. For $N_\T{p}>1$, it approximates
the number of significant spatial channels, whose average normalized strength remains
approximately constant~\cite{Gustafsson+Brick2026}. In contrast, for $N_\T{p}\ll 1$, it determines the normalized strength of the surviving fundamental mode and thereby recovers the Friis transmission formula. The transition $N_\T{p}=1$ defines the characteristic spatial-mode transition distance
$d_\T{b}=\sqrt{A_\T{T}A_\T{R}}/\lambda$. Numerical results for circular
apertures demonstrate the transition and the extension provided by the
mutual-shadow formulation beyond the paraxial regime.

\section{From spatial degrees of freedom to Friis}

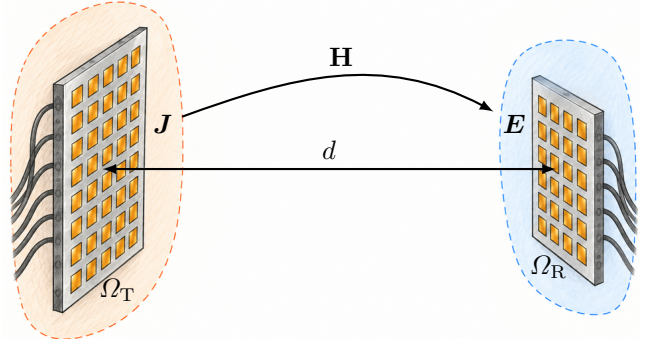
\begin{figure}[t]
\centering
\begin{tikzpicture}
% \node at (0,0) {\includegraphics[width=0.95\linewidth]{figscomp/Array2Array1.png}};
\node at (0,0) {\includegraphics[trim={0cm 0.95mm 0 0.95mm},clip,width=0.95\linewidth]{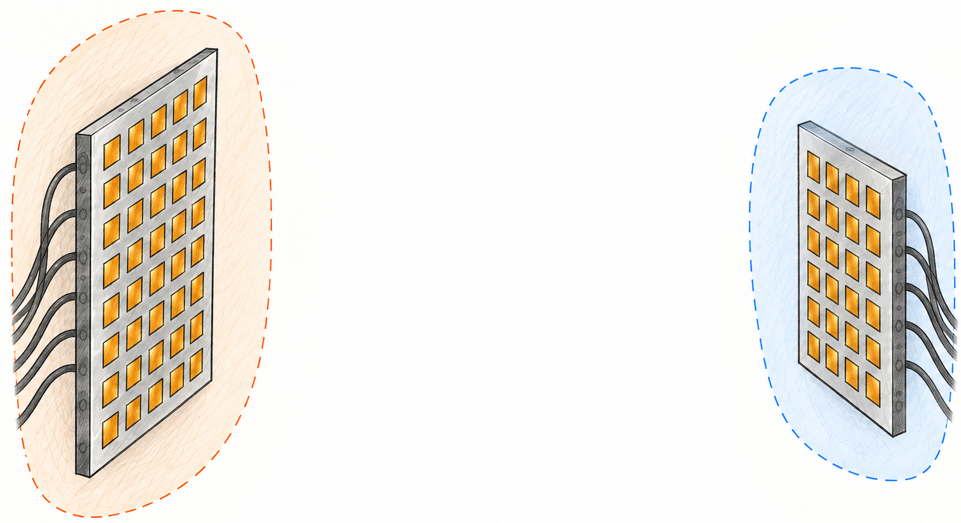}};
\draw[<->,thick] (-2.9,0) -- node[above] {$d$} (3.1,0);
\node at (-2.7,-1.6) {$\regT$};
\node at (-2.1,0.6) (J) {$\V{J}$};
\node at (3,-1.3) {$\regR$};
\node at (2.55,0.6) (E) {$\V{E}$};
\draw[->,thick] (J) [out=20, in=150] to node[above] {$\M{H}$} (E);
\end{tikzpicture}
\caption{Illustration of the considered antenna system, consisting of transmitting and receiving arrays located in $\regT$ and $\regR$, respectively, and separated by a distance $d$. The channel $\M{H}$ maps the current density $\V{J}$ in the transmitting region $\regT$ to the electric field $\V{E}$ in the receiving region $\regR$.}
    \label{fig:ArrayChannel}
\end{figure}

Consider the propagation channel between two finite antenna apertures
$\regT$ and $\regR$, see Fig.~\ref{fig:ArrayChannel}. For electrically large regions, the NDoF per
polarization is asymptotically given by~\cite{Gustafsson2025c}
\begin{equation}
    \Na = \frac{A_\T{TR}}{\lambda^2},
    \label{eq:Na}
\end{equation}
where $A_\T{TR}$ denotes the total mutual shadow (or view) area between the
two apertures. This is the dominant term in the high-frequency asymptotic expansion of the NDoF $N_\T{DoF}=\Na+o(\lambda^{-2})$. A further contribution, associated with the low-frequency (or, similarly, large-distance) regime, gives one DoF per polarization. Including this term extends the approximation's range of validity and gives $N_\T{DoF}\approx\Na+1$~\cite{Gustafsson2025c}. The scalar Green's function considered here yields the DoF per polarization, which is sufficient for the present analysis. Dyadic Green's functions can be incorporated when polarization effects are of interest for specific antenna configurations~\cite{Yuan+etal2022}.

If every point of either aperture is visible from every point
of the other aperture, the mutual shadow area can be expressed as
\cite{Gustafsson2025c,Brick+etal2026}
\begin{equation}
A_\T{TR} =
\int_{\regT}\int_{\regR}
\frac{|\UV{n}_\T{T}\cdot\wUV{R}|\, |\UV{n}_\T{R}\cdot\wUV{R}|}{|\V{R}|^2}
\diffS_\T{R}\diffS_\T{T},
\label{eq:ShadowArea}
\end{equation}
where $\V{R}=\V{r}_\T{R}-\V{r}_\T{T}$ and
$\wUV{R}=\V{R}/|\V{R}|$.

When the separation $d$ is sufficiently large compared with the aperture
dimensions, $|\V{R}|\approx d$ and $\wUV{R}$ is approximately constant over
the two apertures. Equation~\eqref{eq:ShadowArea} then reduces to
\begin{equation}
A_\T{TR} \approx \frac{1}{d^2}
\int_{\regT} |\UV{n}_\T{T}\cdot\wUV{R}|\diffS_\T{T}
\int_{\regR} |\UV{n}_\T{R}\cdot\wUV{R}|\diffS_\T{R}
=
\frac{A_\T{T}A_\T{R}}{d^2},
\label{eq:ShadowAreaParaxial}
\end{equation}
where $A_\T{T}$ and $A_\T{R}$ are the projected areas of the transmitting
and receiving apertures, respectively. Hence,
\begin{equation}
    \Na\approx N_\T{p}
    =
    \frac{A_\T{T}A_\T{R}}{\lambda^2d^2},
    \label{eq:NDoFParaxial}
\end{equation}
which agrees with the established paraxial result for planar apertures
\cite{Miller2019}.

The decrease of $N_\T{p}$ with distance does not imply a corresponding
decrease in the average strength of the supported spatial channels. For the
scalar free-space Green's function and parallel planar apertures, the
Hilbert--Schmidt norm of the propagation operator gives~\cite{Miller2019}
\begin{equation}
\sum_n \sigma_n^2
=\int_{\regT}\int_{\regR}
\frac{\diffS_\T{R}\diffS_\T{T}}{(4\pi)^2|\V{R}|^2}
\approx
\frac{A_\T{T}A_\T{R}}{(4\pi)^2d^2},
\label{eq:paraxial}
\end{equation}
where $\sigma_n=\svd(\M{H})$ are the singular values of the propagation
operator~\cite{Gustafsson+Brick2026}. Here $\M{H}$ denotes the scalar free-space propagation operator from current densities on $\regT$ to fields on $\regR$, see Fig.~\ref{fig:ArrayChannel}. Introducing the normalized singular values
$\sigmat_n=4\pi\sigma_n/\lambda$ and combining
\eqref{eq:NDoFParaxial} and~\eqref{eq:paraxial} gives the central relation~\cite{Gustafsson+Brick2026}
\begin{equation}
    \sum_n\sigmat_n^2
    \approx N_\T{p}.
    \label{eq:SumRule}
\end{equation}

Under the asymptotic assumption that approximately $N_\T{p}$ singular values contribute significantly, their average normalized strength is
\begin{equation}
    \frac{1}{N_\T{p}}
    \sum_n\sigmat_n^2
    \approx 1.
    \label{eq:AverageStrength}
\end{equation}
Thus, increasing separation in the multimode regime primarily removes
spatial channels rather than reducing their average normalized strength.
The multimode contribution becomes comparable to the single far-field mode when $\Na\sim 1$. We therefore define the characteristic DoF-transition distance $d_\T b$ through 
\begin{equation}
N_\T{a}(d_\T{b})=1.
\label{eq:breakDistanceNa}
\end{equation}
Using~\eqref{eq:NDoFParaxial} gives the paraxial estimate
\begin{equation}
d_\T{b}
\approx\frac{\sqrt{A_\T{T}A_\T{R}}}{\lambda}.
\label{eq:breakDistance}
\end{equation}

Unlike the conventional Fraunhofer distance and Fresnel-number criteria, which characterize wavefront and diffraction behavior, $d_\T{b}$ is a link-dependent spatial-mode measure. It depends on the projected areas of both apertures and identifies the transition from multimode to effectively single-mode propagation. In contrast, the Fraunhofer distance $d_\T{f}=2D^2/\lambda$, where $D$ is the maximum dimension of an individual aperture, characterizes the validity of the far-field wavefront approximation~\cite{Balanis2005}. The two distances therefore depend on different geometrical measures and are not related by a universal factor. For identical square apertures of side length $L$, $d_\T{b}=L^2/\lambda$ and $d_\T{f}=4L^2/\lambda$, whereas for circular apertures of radius $a$, $d_\T{b}=\pi a^2/\lambda$ and $d_\T{f}=8a^2/\lambda$. An aperture may therefore satisfy a conventional far-field criterion while the link criterion depends also on the receiving aperture, and vice versa.

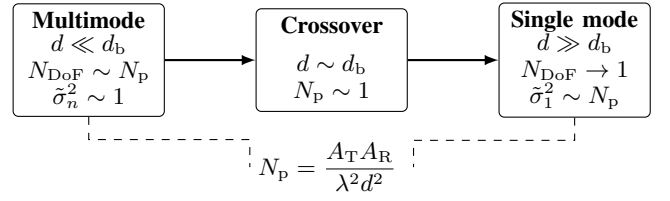
\begin{figure}[t]
\centering
\begin{tikzpicture}[
    font=\small,
    >=latex,
    box/.style={
        draw,
        rounded corners=2pt,
        align=center,
        minimum width=2.0cm,
        minimum height=1.35cm,
        inner sep=3pt
    },
    arrow/.style={->, thick}
]

\node[box] (a) {
    \textbf{Multimode}\\[-0.5mm]
    $d\ll d_{\rm b}$\\
    $N_{\T{DoF}}\sim N_{\T p}$\\
    $\sigmat_n^2\sim1$
};

\node[box, right=12mm of a] (b) {
    \textbf{Crossover}\\[1.5mm]
    $d\sim d_{\rm b}$\\
    $N_{\rm p}\sim1$
};

\node[box, right=12mm of b] (c) {
    \textbf{Single mode}\\[-0.5mm]
    $d\gg d_{\rm b}$\\
    $N_{\T{DoF}}\rightarrow1$\\
    $\sigmat_1^2\sim N_{\T p}$
};

\draw[arrow] (a) -- (b);
\draw[arrow] (b) -- (c);

\node[
    below=3mm of b,
    align=center,
    inner sep=3pt
] (common) {
    $\displaystyle
    N_{\rm p}
    =\frac{A_{\rm T}A_{\rm R}}{\lambda^2d^2}
    $
};

\draw[dashed] (a.south) -- ++(0,-3mm) -| (common.west);
\draw[dashed] (c.south) -- ++(0,-3mm) -| (common.east);

\end{tikzpicture}
\caption{
The common aperture-coupling parameter $N_{\T p}$ changes its physical interpretation across the spatial-mode transition: it approximately
counts the number of significant channels in the multimode regime~\eqref{eq:NDoFParaxial} and determines the dominant-channel strength~\eqref{eq:SVFriis} in the single-mode regime.
}
\label{fig:DoF_Friis_transition}
\end{figure}

For $d$ beyond the transition region around $d_\T{b}$, the channel becomes effectively single mode per
polarization. The sum in~\eqref{eq:SumRule} is then dominated by its
largest singular value,
\begin{equation}
    \sigmat_1^2
    \approx N_\T{p}
    =\frac{A_\T{T}A_\T{R}}{\lambda^2d^2}.
    \label{eq:SVFriis}
\end{equation}
The physical interpretation of $N_\T{p}$ therefore changes: for $N_\T{p}\gg 1$ it approximately counts the number of spatial channels, whereas
for $N_\T{p}\ll 1$ it describes the normalized strength of the surviving
fundamental channel. Fig.~\ref{fig:DoF_Friis_transition} summarizes these regions. 

With the adopted normalization, the projected geometrical areas coincide with the effective areas for polarization-matched, unity-efficiency apertures. The Friis power-transfer ratio~\cite{Schelkunoff+Friis1952,Balanis2005} can then be written as
\begin{equation}
    \frac{P_\T{R}}{P_\T{T}}
    =\frac{A_\T{T}A_\T{R}}{\lambda^2d^2}.
    \label{eq:Friis}
\end{equation}
Consequently,
\begin{equation}
    \frac{P_\T{R}}{P_\T{T}}
    \approx \sigmat_1^2
    \approx N_\T{p},
    \qquad d\gtrsim d_\T{b}.    
    \label{eq:FriisConnection}
\end{equation}
The Friis transmission formula can thus be interpreted as the single-mode
continuation of the same geometrical quantity that determines the spatial
DoF in the multimode regime. This interpretation also extends naturally to the multimode region $d<d_\T{b}$, where, under equal power allocation among the available spatial channels, the same coupling parameter yields a Friis-like expression for the total received power.

%%%%%%%%%%%%%%%%%%%%%%%%%%%%%%
\section{Numerical example: Circular apertures}

To illustrate the transition from multimode near-field propagation to
single-mode far-field transmission, consider two parallel circular apertures
of radii $a$ and $b$, separated by a distance $d$. We first consider equal
apertures with $a=b=10\lambda$ and subsequently investigate the dependence
on electrical size and aperture ratio.

\begin{figure}
    \centering
\includegraphics[width=\linewidth]{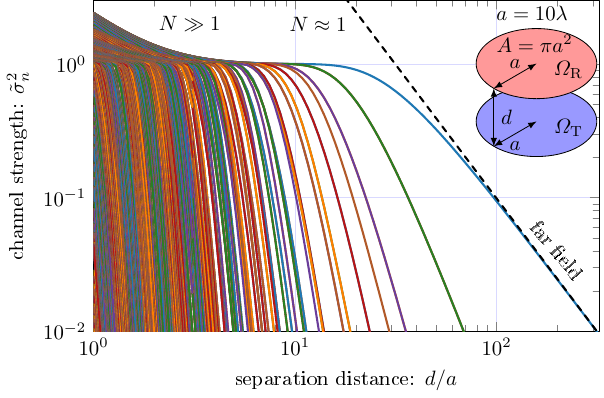}
\caption{Normalized squared channel singular values $\sigmat_n^2$ between
two circular apertures with radius $a=10\lambda$ as a function of
separation distance $d$. The dashed curve denotes the single-mode
far-field prediction~\eqref{eq:SVFriis}. The channel transitions to an effectively single-mode regime around $d_\T{b}=10\pi a$.}
    \label{fig:EigsDiscsFriis}
\end{figure}

The normalized squared singular values $\sigmat_n^2$ for two equal circular apertures with $a=10\lambda$ are shown in Fig.~\ref{fig:EigsDiscsFriis}. At short separations, the channel supports many spatial modes with normalized strengths of order unity. As the separation increases, the higher-order modes progressively weaken and cease to contribute significantly, while the strongest mode remains approximately constant over the multimode region. Around the beamforming distance $d=d_\T{b}=10\pi a$, the channel transitions to an effectively single-mode regime. For larger separations, the dominant mode exhibits the quadratic distance dependence predicted by~\eqref{eq:SVFriis}. The dashed curve therefore provides a direct connection between the numerical singular-value spectrum and the Friis transmission limit.

\begin{figure}
    \centering
    \includegraphics[width=\linewidth]{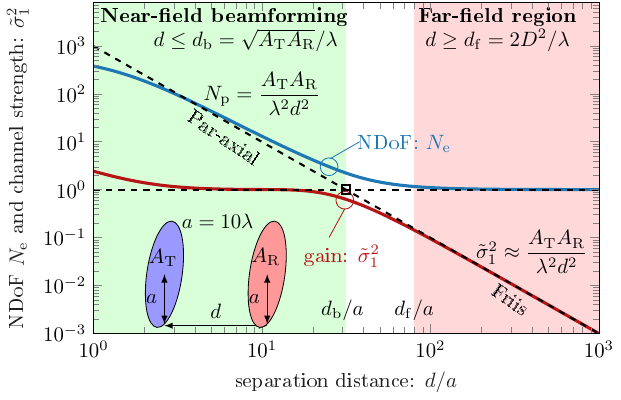}
\caption{Effective NDoF $N_\T{e}$ and dominant normalized channel strength
$\sigmat_1^2$ for two circular apertures with radius $a=10\lambda$.
The near-field beamforming region supports approximately $\Np$ spatial
channels with approximately constant average normalized strength,
whereas beyond $d_\T{b}$ the channel is effectively single mode and
$\sigmat_1^2$ follows the Friis $d^{-2}$ dependence, see also Fig.~\ref{fig:DoF_Friis_transition}.}
    \label{fig:DoF_Friis_Discs}
\end{figure}

The effective NDoF is characterized by~\cite{Gustafsson+Brick2026}
\begin{equation}
    N_\T{e}
    =
    \frac{\left(\sum_n\sigmat_n^2\right)^2}
    {\sum_n\sigmat_n^4}.
    \label{eq:Ne}
\end{equation}
This measure approaches the actual number of equally strong spatial channels and gives $N_\T{e}=1$ for a rank-one channel. Figure~\ref{fig:DoF_Friis_Discs} compares $N_\T{e}$ with the paraxial NDoF $\Np$ and the dominant normalized
channel strength $\sigmat_1^2$. In the multimode regime, $N_\T{e}$ decreases approximately as $d^{-2}$, in agreement with the paraxial prediction, while the dominant channel strength remains of order unity. Around $d=d_\T{b}$, the effective NDoF approaches unity. Beyond this transition, $N_\T{e}\approx1$, whereas the strength of the remaining channel decreases as $d^{-2}$ in accordance with~\eqref{eq:SVFriis}.

Thus, the same distance dependence has two distinct physical interpretations. Below the beamforming distance, increasing separation primarily reduces the number of spatial channels. Above it, the spatial rank has essentially saturated at one and further separation instead reduces the strength of the remaining channel.

\begin{figure}
    \centering
    \includegraphics[width=\linewidth]{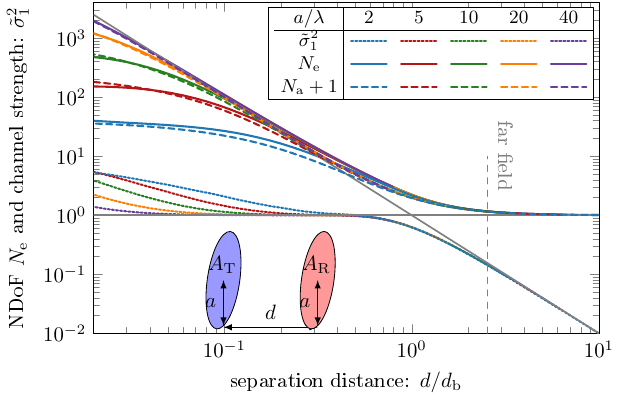}
\caption{Effective NDoF $N_\T{e}$ and dominant normalized channel strength
$\sigmat_1^2$ for equal circular apertures with
$a\in\{2,5,10,20,40\}\lambda$, with separation normalized by the transition distance $d_\T{b}$. The effective NDoF is compared with the
asymptotic mutual-shadow prediction $\Na+1$ and the paraxial approximation
$\Np$.}    \label{fig:DoF_Friis_Discs_d_la}
\end{figure}

The dependence on electrical aperture size is investigated in
Fig.~\ref{fig:DoF_Friis_Discs_d_la}. The results are shown for equal circular
apertures with radii $a\in\{2,5,10,20,40\}\lambda$ and the separation
distance normalized by the corresponding beamforming distance $d_\T{b}$.
The normalization produces a similar transition for the different
electrical sizes, indicating that $d_\T{b}$ provides the relevant distance
scale for the onset of the single-mode regime.

For $d\gtrsim d_\T{b}$, the dominant channel follows the same normalized
quadratic decay for all aperture sizes, consistent with~\eqref{eq:SVFriis}.
For $d\lesssim  d_\T{b}$, several spatial modes contribute and the effective NDoF
increases with decreasing separation. The paraxial approximation describes
the transition region well, while deviations become increasingly visible
at shorter distances where the assumptions leading to~\eqref{eq:NDoFParaxial}
are no longer accurate. The mutual-shadow prediction provides a more
accurate description in this regime.

\begin{figure}
    \centering
    \includegraphics[width=\linewidth]{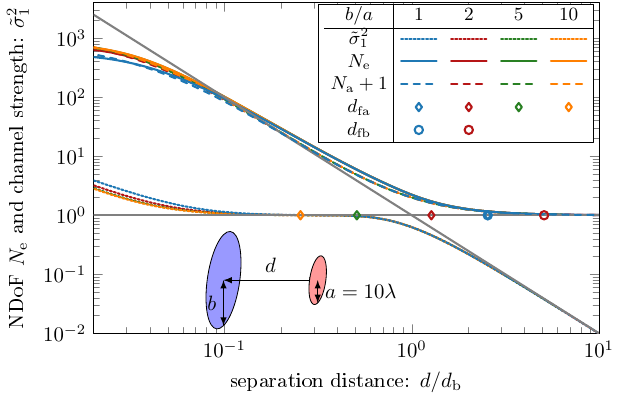}
\caption{Effective NDoF $N_\T{e}$ and dominant normalized channel strength
$\sigmat_1^2$ for a circular transmitting aperture with $a=10\lambda$ and
receiving apertures with $b\in\{1,2,5,10\}a$, with the separation
normalized by the corresponding spatial-mode transition distance $d_\T{b}$. The
effective NDoF is compared with the mutual-shadow prediction $\Na+1$ and
the paraxial approximation $\Np$. The corresponding far-field distances $d_\T{f}$ are indicated by markers.}   \label{fig:DoF_Friis_Discs_d_Radii}
\end{figure}

The dependence on the relative aperture sizes is investigated in
Fig.~\ref{fig:DoF_Friis_Discs_d_Radii} for a transmitting aperture with
radius $a=10\lambda$ and receiving apertures with
$b\in\{1,2,5,10\}a$. The separation is normalized by the
corresponding spatial-mode transition distance~\eqref{eq:breakDistance}. The effective
NDoF and the mutual-shadow prediction exhibit a consistent transition
around $d=d_\T{b}$ over a wide range of aperture ratios. The far-field distances $d_\T{f}$ are indicated by markers. 

The dependence on both aperture sizes is a characteristic feature of the proposed transition distance. In the paraxial regime,
$d_\T{b}=\sqrt{A_\T{T}A_\T{R}}/\lambda$ depends on the product of the
transmitting and receiving projected areas, rather than on either aperture
separately. The numerical results therefore support interpreting
$d_\T{b}$ as a property of the complete propagation channel rather than
as an antenna-specific near-field boundary.

The deviations from the paraxial approximation at short distances are
further illustrated by the mutual-shadow area. For two parallel circular
apertures with radii $a$ and $b\geq a$, the mutual shadow can be evaluated
analytically as~\cite{Gustafsson+Brick2026}
\begin{equation}
    A_\T{TR}  =
    \frac{\pi^2}{2}
    \big(
        \varDelta-
        \sqrt{\varDelta^2-4a^2b^2}
    \big)
     \sim
    \begin{cases}
        \pi^2a^2, & d\ll a\\[1mm]
        \frac{\pi^2a^2b^2}{d^2}, & d\gg b
    \end{cases}   
    % ,
    % \quad
    % \varDelta=a^2+b^2+d^2,
    \label{eq:ShadowA2discs}
\end{equation}
with $\varDelta=a^2+b^2+d^2$.
The corresponding asymptotic NDoF therefore saturates at short distances
and approaches the paraxial $d^{-2}$ dependence at large separation.

For the smaller separations shown in Figs~\ref{fig:DoF_Friis_Discs_d_la} and~\ref{fig:DoF_Friis_Discs_d_Radii},
where the paraxial approximation becomes less accurate, the mutual-shadow
prediction follows the numerical effective NDoF more closely. This
demonstrates the advantage of using the mutual shadow as the underlying
geometrical measure: the simple paraxial expression provides the useful
closed-form beamforming distance, while the full mutual-shadow expression
extends the description into the nonparaxial regime. Inverting~\eqref{eq:ShadowA2discs} for $A_\T{TR}=\lambda^2$ and expanding in the electrically large-aperture limit gives
\begin{equation}
d_\T{b} \approx \frac{\sqrt{A_\T{T}A_\T{R}}}{\lambda}
\left(1-\frac{1}{2\pi}\left(\frac{\lambda^2}{a^2}+\frac{\lambda^2}{b^2}\right)\right).
\end{equation}
The correction is therefore of second order in the small parameters $\lambda/a$ and $\lambda/b$, confirming that the paraxial expression~\eqref{eq:breakDistance} is accurate for $\lambda\ll a,b$.

\section{Conclusion}
The transition from multimode near-field propagation to single-mode far-field transmission can be described by a common geometrical coupling parameter. In the paraxial regime, this parameter is $
N_\T{p}=A_\T{T}A_\T{R}/(\lambda^2d^2)$, and is also proportional to the total normalized strength of the propagation channel. For \(N_\T{p}\gg1\), it primarily determines the number of significant spatial modes, whose average normalized strength remains of order unity. For \(N_\T{p}\ll1\), the channel is effectively single mode and the same parameter instead determines the strength of the dominant mode, recovering the Friis transmission formula for ideal polarization-matched apertures.

The crossover \(N_\T{p}\sim1\) defines the characteristic spatial-mode transition distance
$d_\T{b}=\sqrt{A_\T{T}A_\T{R}}/\lambda$.
Unlike conventional Fraunhofer and Fresnel distances, this transition depends on both apertures and therefore characterizes the complete transmitter--receiver link. The mutual-shadow formulation extends this interpretation beyond the paraxial regime, where it accounts for aperture shape, orientation, and the saturation of the spatial degrees of freedom at short distances.

These results show that spatial multiplexing in the near field and Friis transmission in the far field need not be regarded as separate propagation descriptions. Instead, they represent two limiting manifestations of the same aperture-to-aperture coupling: with increasing separation, the geometrical coupling first determines how many spatial channels can be supported and, after the channel becomes effectively rank one, determines the strength of the remaining channel.

%%%%%%%%%%%%%%%%%%%%%%%%%%%%%%%%%%%%%%%%%%%%%%
%\bibliographystyle{IEEEtran}
%\bibliography{total,bibadd}


\end{document}